\documentclass[english,twocolumn,floats,prl,aps,preprintnumbers,superscriptaddress]{revtex4}
\usepackage[T1]{fontenc}
\usepackage[latin9]{inputenc}
\usepackage{amsmath}
\usepackage{amssymb}
\usepackage{esint}
\usepackage{epstopdf}
\usepackage{graphicx}
\usepackage{float}

\makeatletter
\@ifundefined{textcolor}{}
{%
 \definecolor{BLACK}{gray}{0}
 \definecolor{WHITE}{gray}{1}
 \definecolor{RED}{rgb}{1,0,0}
 \definecolor{GREEN}{rgb}{0,1,0}
 \definecolor{BLUE}{rgb}{0,0,1}
 \definecolor{CYAN}{cmyk}{1,0,0,0}
 \definecolor{MAGENTA}{cmyk}{0,1,0,0}
 \definecolor{YELLOW}{cmyk}{0,0,1,0}
}

\usepackage{dcolumn}
\usepackage{bm}
\usepackage{enumerate}

\makeatother

\usepackage{babel}
\begin{document}

\title{Multiple steady-states in nonequilibrium quantum systems with
  electron-phonon interactions}

\author{Eli Y. Wilner} \affiliation{School of Physics and Astronomy,
  The Sackler Faculty of Exact Sciences, Tel Aviv University,Tel Aviv
  69978,Israel}

\author{Haobin Wang} \affiliation{Department of Chemistry and
  Biochemistry, New Mexico State University, Las Cruces, NM 88003, USA}

\author{Guy Cohen} \affiliation{Department of Physics and Chemistry,
  Columbia University, New York, New York 10027, USA}

\author{Michael Thoss} \affiliation{Institute for Theoretical Physics
  and Interdisciplinary Center for Molecular Materials,
  Friedrich-Alexander-Universit{\"a}t Erlangen-N{\"u}rnberg,
  Staudtstr. 7/B2, D-991058 Erlangen, Germany}

\author{Eran Rabani} \affiliation{School of Chemistry, The Sackler
  Faculty of Exact Sciences, Tel Aviv University,Tel Aviv
  69978,Israel}

\date{\today}
\begin{abstract}
The existence of more than one steady-state in a many-body quantum
system driven out-of-equilibrium has been a matter of debate, both in
the context of simple impurity models and in the case of inelastic
tunneling channels.  In this Letter, we combine a reduced density
matrix formalism with the multilayer multiconfiguration time-dependent
Hartree method to address this problem.  This allows to obtain a
converged numerical solution of the nonequilibrium dynamics.
Considering a generic model for quantum transport through a quantum
dot with electron-phonon interaction, we prove that a unique
steady-state exists regardless of the initial electronic preparation
of the quantum dot consistent with the converged numerical results.
However, a bistability can be observed for different initial phonon
preparations.  The effects of the phonon frequency and strength of the
electron-phonon couplings on the relaxation to steady-state and on the
emergence of bistability is discussed.
\end{abstract}

\maketitle

The existence of a unique steady-state in strongly correlated quantum
systems out-of-equilibrium has been a subject of great interest and
controversy.  For the case of the Anderson impurity model, it has been
argued using the Bethe ansatz that a single steady-state solution
exists~\cite{Andrei2006}. However, recent calculations of the
nonequilibrium current based on time-dependent density functional
theory seem to indicate that at long times the system reaches a
dynamical state characterized by correlation-induced current
oscillations~\cite{Gr2010}.  Similarly, questions regarding
hysteresis, bistability and the dependence of the steady-state current
on the initial occupation have been raised in the context of inelastic
transport through nanoscale quantum
dots~\cite{galperin_hysteresis_2004,galperin_non-linear_2008,Bratkovsky2007,Bratkovsky2009,Kosov2011,albrecht_bistability_2012,Komnik2013}.

Addressing the issue of a unique steady-state is a challenging task
for theory, as systems exhibiting bistability involve strong
electron-electron or electron-phonon correlations. Under these
conditions, an exact solution is unavailable, and one has to resort to
approximate methods or to numerical techniques.  The former are based
on either a mean-field approximation or on a perturbative scheme,
where the inclusion of higher order corrections is not always clear or
systematic, and thus may lead to questionable results.  Numerically
brute-force approaches, such as time-dependent numerical
renormalization-group
techniques~\cite{anders_real-time_2005,schmitteckert_nonequilibrium_2004,white_density_1992},
iterative~\cite{weiss_iterative_2008,eckel_comparative_2010,Segal10}
or
stochastic~\cite{muhlbacher_real-time_2008,werner_diagrammatic_2009,schiro_real-time_2009,werner_weak-coupling_2010}
diagrammatic methods, and wave function based
approaches~\cite{wang_numerically_2011}, have been very fruitful, but
are limited in the range of parameters and timescales that can be
studied.

In this Letter, we address the problem of a unique steady-state in a
system with electron-phonon couplings.  To study the nonequilibrium
dynamics, we develop an approach based on a reduced density matrix
(RDM) formalism, which requires as input a short-lived memory
kernel~\cite{cohen_memory_2011}.  We show that if a steady-state
exists then it must be unique, regardless of the initial electronic
preparation.  However, a bistability can develop for different initial
phonon states. The relaxation to steady-state and the appearance of
the bistability depends on the phonon frequency and the strength of
the electron-phonon couplings. To illustrate this, we combine the
approach with the multilayer multiconfiguration time-dependent Hartree
method (ML-MCTDH)~\cite{Thoss03} to numerically converge the memory
kernel at short times until it decays, and infer from it the dynamics
of the system at all times and the approach to steady-state.

We consider a generic model for charge transport through a quantum dot
with electron-phonon interaction.  The model is described by the
Hamiltonian $H=H_{S}+H_{B}+V_{SB}$, where
$H_{S}=\varepsilon_{d}d^{\dagger}d$ is the system (quantum dot)
Hamiltonian with creation/annihilation fermionic operators
$d^{\dagger}$/$d$ and energy $\varepsilon_d$, $H_{B}=H_{l}+H_{\mbox {
    ph}}$ where $H_{l}=\sum_{k\in
  L,R}\varepsilon_{k}a_{k}^{\dagger}a_{k}$ represents the
noninteracting leads Hamiltonian with fermionic creation/annihilation
operators $a_{k}^{\dagger}$/$a_{k}$, and $H_{\mbox
  ph}=\underset{\alpha} {\sum}\omega_{\alpha}
\left(b_{\alpha}^{\dagger}b_{\alpha}+ \frac{1}{2}\right)$ represents
the phonon bath with creation/annihilation bosonic operators
$b_{\alpha}^{\dagger}$/$b_{\alpha}$ for phonon mode $\alpha$ with
energy $\omega_{\alpha}$. The coupling between the system and the
baths is given by $V_{SB}=V_{l}+V_{\mbox ph}$ where
$V_l=\underset{k\in L,R}{\sum}\left(t_{k}da_{k}^{\dagger} +
t_{k}^{*}a_{k}d^{\dagger}\right)$ is the coupling between the system
and the leads with couplings strength $t_k$, and $V_{\mbox
  ph}=d^{\dagger}d\underset{\alpha}{\sum}
M_{\alpha}\left(b_{\alpha}^{\dagger} + b_{\alpha}\right)$ is the
couplings between the system and the phonon bath, where $M_{\alpha}$
is the electron-phonon couplings to mode $\alpha$.

The coupling strengths were determined by various spectral functions.
The dot-leads coupling terms were determined from a spectral function
of the form $\Gamma_{L,R}(\varepsilon) = 2\pi \sum_{k \in L,R}
|t_{k}|^2 \delta(\varepsilon-\varepsilon_{k})$, where a tight-binding
form was employed: $\Gamma_{L,R}\left(\varepsilon\right) =
\frac{a^{2}} {b^{2}} \sqrt{4b^{2} - (\varepsilon-\mu_{L,R})^{2}}$ with
$a=0.2\mbox{eV}$ and $b=1\mbox{eV}$. $\mu_{L,R}$ is the chemical
potential of the left (L) or right (R) lead, respectively.  Similarly,
the electron-phonon couplings were determined from a phonon spectral
function $J(\omega) = \frac{\pi}{2} \sum_{\alpha}
\frac{M_{\alpha}^{2}}{\omega_{\alpha}}
\delta(\omega-\omega_{\alpha})$, where $J\left(\omega\right) =
\frac{\pi}{2} \eta \omega e^{-\frac{\omega}{\omega_{c}}}$ is taken to
be of Ohmic form. The dimensionless Kondo parameter, $\eta =
\frac{\lambda} {2\omega_c}$, determines the overall strength of the
electron-phonon couplings where $\lambda=\sum \frac{M_{\alpha}^{2}}
{2\omega_{\alpha}} = \frac{1}{\pi}\int \frac{d\omega }{\omega}
J(\omega)$ is the reorganization energy (or polaron shift) and
$\omega_c$ is the characteristic phonon bath frequency.

Following the derivation outlined in
Ref. \onlinecite{cohen_memory_2011} for the Anderson impurity model,
the equation of motion for the RDM, $\sigma(t)=Tr_B\{\rho(t)\}$, is
given by
\begin{equation}
i\hbar\frac{\partial}{\partial t}\sigma\left(t\right) =
\mathcal{L}_{S}\sigma\left(t\right) + \vartheta\left(t\right) -
\frac{i}{\hbar} \int_{0}^{t}d\tau\kappa\left(\tau\right)\sigma
\left(t-\tau\right)
\label{eq:sigma(t)}
\end{equation}
where $\mathcal{L}_{S}=[H_{S},\cdots]$ is the system's Liouvillian,
$Tr_B\{\cdots\}$ is a trace over the baths degrees of freedom (leads
and phonon baths) and $\rho(t)$ is the full density matrix.  In the
above,
\begin{equation}
\vartheta\left(t\right) = Tr_{B}\left\{ \mathcal{L}_{V}
e^{-\frac{i}{\hbar}Q\mathcal{L}t}Q\rho\left(0\right)\right\}
\label{eq:theta(t)}
\end{equation}
depends on the choice of initial conditions and vanishes for an
uncorrelated initial state (which is the case discussed below), {\em
  i.e.} when $\rho(0)= \sigma(0) \otimes \rho_B(0)$, where $\sigma(0)$
and $\rho_B(0)$ are the system and baths initial density matrices,
respectively, and $\mathcal{L}_{v}=[V_{SB},\cdots]$. The memory
kernel, which describes the non-Markovian dependency of the time
propagation of the system, is given by
\begin{equation}
\kappa\left(t\right) = Tr_{B}\left\{ \mathcal{L}_{V}
e^{-\frac{i}{\hbar}Q\mathcal{L}t} Q\mathcal{L}\rho_{B}\right\}
\label{eq:memory-kernel}
\end{equation}
where $Q=1-P$, $P=\rho_B(0) Tr_B\{\cdots\}$ and
$\mathcal{L}=[H,\cdots]$.

To obtain $\sigma(t)$, one requires as input the super-matrix of the
memory kernel, which can be expressed in terms of a Volterra equation
of the second type, removing the complexity of the projected dynamics
of Eq.~(\ref{eq:memory-kernel}):
\begin{equation}
\kappa\left(t\right) = i\hbar\dot{\Phi}\left(t\right) -
\Phi\left(t\right) \mathcal{L}_{S} + \frac{i}{\hbar}
\int_{0}^{t}d\tau\Phi\left(t-\tau\right) \kappa \left(\tau\right)
\label{eq:volterra}
\end{equation}
with
\begin{equation}
\Phi\left(t\right)=Tr_{B}\left\{ \mathcal{L}_{V}e^{-\frac{i}{\hbar}\mathcal{L}t}\rho_{B}\right\}.
\label{eq:Phi(t)}
\end{equation}
Evaluating the super-matrix $\Phi(t)$ requires a tedious calculation
similar to that carried out in
Ref.~\onlinecite{cohen_memory_2011}. The diagonal matrix elements
$\Phi_{ii,mm}(t)$ are given by $\frac{2}{\hbar}\Im\left\{
\varphi_{mm}(t)\right\}$, where
\begin{equation}
\varphi_{mn}(t) = Tr_{B}\left\{ \rho_{B}\left\langle n\right|
\underset{k} {\sum}t_{k}d(t)a_{k}^{\dagger}(t)\left|m\right\rangle
\right\}.
\label{eq:phi(t)}
\end{equation}
The indices $i$, $j$, $m$, and $n$ can take the values $1$ or $0$,
corresponding to an occupied or an unoccupied dot, respectively.  The
off-diagonal elements of $\Phi_{ij,nm}(t)$ are given by
$\delta_{j1}\delta_{i0}\psi_{mn}(t)+\delta_{j0}\delta_{i1}\psi_{nm}^{*}(t)$,
where
\begin{equation}
\begin{split}  
\psi_{mn}(t) &= Tr_{B}\left\{ \rho_{B}\left\langle n\right|\underset{k}
    {\sum}t_{k}a_{k}^{\dagger}(t) \right. \\ &- \left.
    d^{\dagger}(t)\underset{\alpha}{\sum}
    M_{\alpha} \left(b_{\alpha}^{\dagger}(t) + b_{\alpha}(t)\right)
    \left|m\right\rangle \right\} .
\end{split}    
\label{eq:psi(t)}
\end{equation}

Since both the operator $d(t) a_{k}^{\dagger}(t)$ appearing in the
equation for $\varphi_{mn}(t)$ and the full Hamiltonian conserve the
total particle number, $\varphi_{mn}(t)$ is nonzero only for $m=n$, in
which case it has a simple physical interpretation as the time
derivative of the dot population. $\varphi_{mm}(t)$ can be expressed
in terms of the sum of the left and right currents $e
\varphi_{mm}(t)=I^L_m(t)+I^R_m(t)$, where:
\begin{equation}
I^{L,R}_m(t)=-\frac{2e}{\hbar} \Im \underset {k\in L,R} {\sum} t_{k}
\langle m| d(t) a_{k}^{\dagger}(t)|m\rangle,
\label{eq:current}
\end{equation}
and $e$ is the electron charge.  In addition, one can show that
$\psi_{mn}(t)$ is nonzero only for $m\neq n$. As a result, the
populations and coherences of $\sigma(t)$ are
decoupled~\cite{cohen_memory_2011}, and if one is interested in the
behavior of the populations alone, only $\varphi_{mm}(t)$ is
required. The resulting equations of motion for the diagonal elements
of $\sigma(t)$ are given by:
\begin{equation}
\frac{\partial}{\partial t}\sigma_{ii}\left(t\right) =
-\frac{1}{\hbar^{2}}\underset{m} {\sum}\int_{0}^{t}d\tau
\kappa_{ii,mm}\left(\tau\right)\sigma_{mm}\left(t-\tau\right).
\label{eq:populations}
\end{equation}

If $\sigma(t)$ has a steady-state solution as $t\rightarrow\infty$,
then $\frac{\partial}{\partial t}\sigma_{ii}\left(t\right)=0$, and
$\underset{m} {\sum} \int_{0}^{\infty}d\tau
\kappa_{ii,mm}\left(\tau\right)\sigma_{mm}\left(t-\tau\right)=0$ can
be replaced by a linear set of equations given by:
\begin{equation}
\underset{m} {\sum} \mathcal{K}_{im} \sigma_{m}=0,
\label{eq:Ksigma}
\end{equation}
where $\mathcal{K}_{im}=\frac{1}{\hbar^{2}} \int_{0}^{\infty} d\tau
\kappa_{ii,mm} \left(\tau\right)$ and $\sigma_{m} \equiv
\sigma_{mm}(t\rightarrow\infty)$. The matrix $\mathcal{K}$ in
Eq.~(\ref{eq:Ksigma}) has two eigenvalues. The first is
$\lambda_{0}=0$ corresponding to
$\sigma_{00}=\frac{\mathcal{K}_{00}}{\mathcal{K}_{00} +
  \mathcal{K}_{11}}$ and $\sigma_{11}=1-\sigma_{00} =
\frac{\mathcal{K}_{11}}{\mathcal{K}_{00} + \mathcal{K}_{11}}$.  For a
physical steady-state solution, both $\mathcal{K}_{00}$ and
$\mathcal{K}_{11}$ must share the same sign.  Otherwise, the diagonal
elements of $\sigma$ cannot both be positive. The other eigenvalue is
$\lambda_{1}=\mathcal{K}_{00} + \mathcal{K}_{11}$ corresponding to
$\sigma_{00}=-\sigma_{11}=1$, which can never represent a physical
solution.  Furthermore, since the steady-state depends only on the
value of $\mathcal{K}_{00}$ and $\mathcal{K}_{11}$ and since both are
independent of the initial dot population, the steady-state is
independent of the initial preparation of the dot occupation and
therefore, is unique.

The steady-state solution is unique with respect to the electronic
initial preparation.  This is a strong statement by itself, but it
does not rule out bistability for different initial phonon
preparations.  To address this question, we combined the formalism
described above for the RDM with the ML-MCTDH approach in second
quantized representation
(SQR)~\cite{Wang2009,wang_numerically_2011}. The ML-MCTDH-SQR provides
a tool to compute the currents in Eq.(\ref{eq:current}) numerically
exactly.  The kernel $\kappa(t)$ is then obtained by numerically
solving Eq.~(\ref{eq:volterra}). In comparison, for most model
parameters studied in this work, it is practically impossible to
obtain converged values for the RDM directly from the ML-MCTDH-SQR,
since the time to reach a steady-state solution is significantly
longer than the maximum simulation time reachable by the
ML-MCTDH-SQR. However, since the memory kernel decays on much shorter
timescales compared to the RDM itself~\cite{cohen_memory_2011}, it is
rather straightforward to calculate it using the ML-MCTDH-SQR and then
solve Eq.~(\ref{eq:populations}) for the RDM.

To characterize the population dynamics, we start with a factorized
initial condition of the form $\rho\left(0\right)=\sigma\left(0\right)
\otimes\rho_{\mbox{ph}}\left(0\right) \otimes \rho_{\mbox {leads}}(0)$,
where $\sigma\left(0\right)$ determines whether the electronic level
is initially occupied/unoccupied, $\rho_{\mbox{ph}}\left(0\right) =
\exp\left[-\beta\left\{ \underset{\alpha} {\sum} \omega_{\alpha}
  \left(b_{\alpha}^{\dagger} b_{\alpha} + \frac{1}{2}\right) +
  \underset{\alpha} {\sum} \delta_{\alpha} \left(b_{\alpha}^{\dagger}
  + b_{\alpha}\right)\right\} \right]$ represents the initial density
matrix of the phonon bath.  Hereby two values of the parameters
$\delta_{\alpha}$ are considered: $\delta_{\alpha} = 0$ (corresponding
to a phonon initial state equilibrated with an unoccupied dot) and
$\delta_{\alpha}= \sqrt{2\omega_{\alpha} \lambda}$ (corresponding to
phonons equilibrated to an occupied dot).  $\rho_{\mbox {leads}}(0) =
\exp \left[-\beta \left(\underset{k\in L} {\sum }\left(\varepsilon_{k}
  - \mu_{L}\right) a_{k}^{\dagger}a_{k} + \underset{k\in R}{\sum}
  \left(\varepsilon_{k} - \mu_{R}\right) a_{k}^{\dagger} a_{k}\right)
  \right]$ determines the initial density matrix for the leads. In the
above $\beta=\frac{1}{k_{B}T}$ is the inverse temperature.  In all
results shown below we take $T=0$ and $\mu_L=-\mu_R=0.05$eV.

\begin{figure}[t]
\includegraphics[width=8cm]{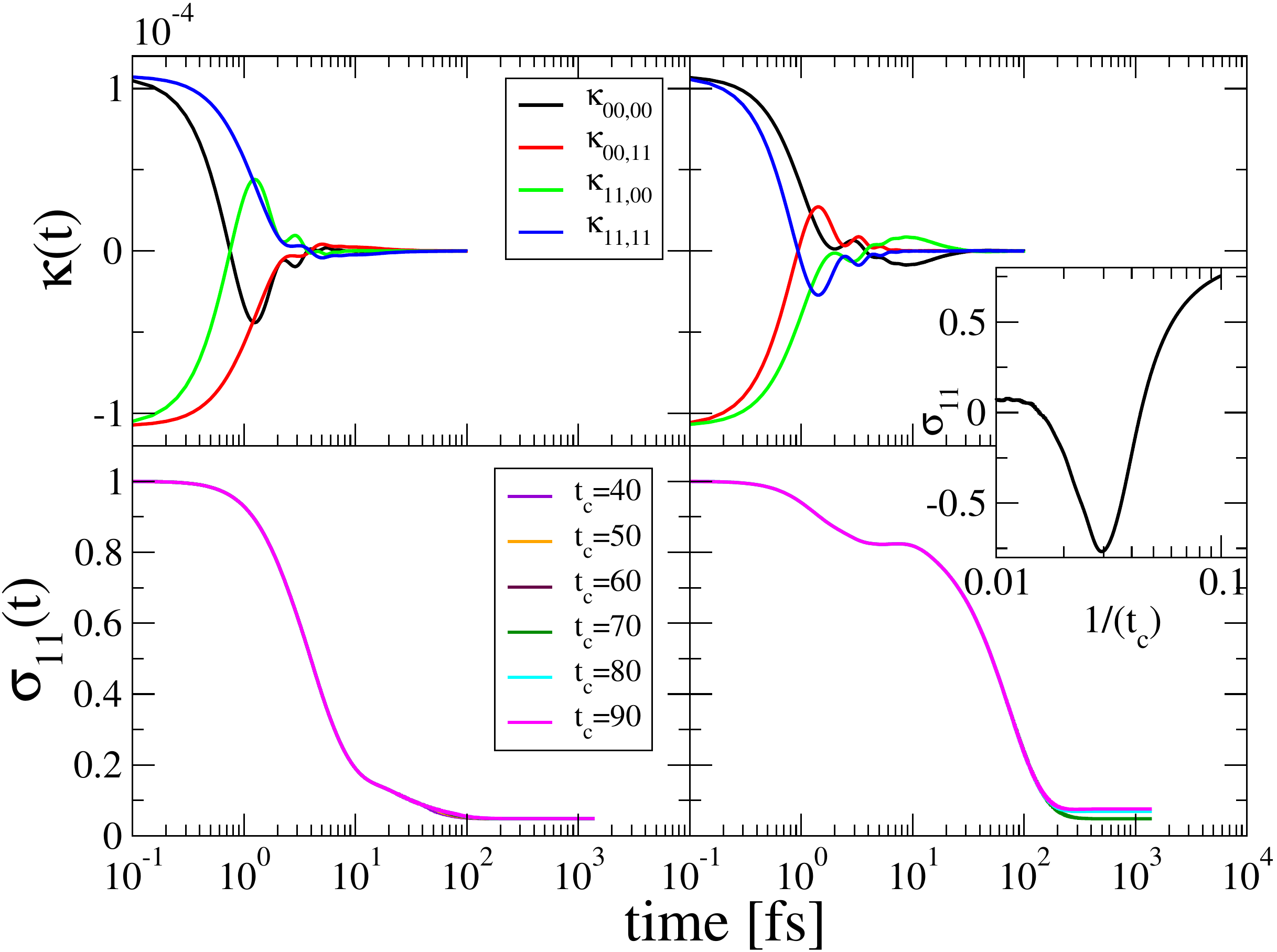}
\caption{The occupation of the quantum dot $\sigma_{11}(t)$ (lower
  panels) for different cutoff times and the nonzero elements of the
  memory kernel (upper panels) for $\delta_{\alpha}=0$ (left panels)
  and $\delta_{\alpha}=\sqrt{2\omega_{\alpha}\lambda}$ (right
  panels). The inset shows the steady-state value of $\sigma_{11}$ for
  the case of $\delta_{\alpha}=\sqrt{2\omega_{\alpha}\lambda}$ as a
  function of $1/t_c$. The model parameters used are:
  $\varepsilon_d=0.5$eV, $\omega_c=500\mbox{cm}^{-1}$, and
  $\lambda=3000\mbox{cm}^{-1}$.}
\label{fig:cutoff_dep}
\end{figure}

In Fig.~\ref{fig:cutoff_dep} we plot the time evolution of
$\sigma_{11}(t)$ (lower panels) and the corresponding nonzero elements
of the memory kernel (upper panels), for two different initial
vibrational preparations.  We show the time evolution of
$\sigma_{11}(t)$ for different values of the cutoff time $t_c$ at
which we assume that the memory kernel has essentially decayed to
zero, such that it can be safely truncated.  For $\delta_{\alpha}=0$,
it is safe to truncate the memory kernel at $t_c>30\mbox{fs}$ while
$\delta_{\alpha}=\sqrt{2\omega_{\alpha}\lambda}$ requires a larger
cutoff time of $t_c>80\mbox{fs}$.  Comparing the time it takes for the
memory kernel to decay (upper panels of Fig.~\ref{fig:cutoff_dep})
with the time taken by the RDM to reach steady-state (corresponding
lower panels of Fig.~\ref{fig:cutoff_dep}), it is clear that the
latter is slower by nearly an order of magnitude and in some cases
even more.  Since the calculation of the memory kernel using the
ML-MCTDH-SQR is by far the most time consuming portion of the
calculation, the combination with the RDM formalism provides a
significant saving, and more importantly extends the ML-MCTDH-SQR
approach to regimes inaccessible by direct application.

The inset of Fig.~\ref{fig:cutoff_dep} shows the steady-state value of
$\sigma_{11}$ as a function of $1/t_c$ for
$\delta_{\alpha}=\sqrt{2\omega_{\alpha}\lambda}$.  For large values of
$1/t_c$ (short cutoff times) we find that the formalism may lead to
unphysical situations in which $\sigma_{11}$ becomes negative. Of
course, this is expected, since only when the memory kernel has
decayed to zero does the cutoff approximation provide a physical
meaningful solution.  As $1/t_c$ decreases $\sigma_{11}$ converges and
approaches a plateau value.  In the present case of parameters, the
steady-state value of $\sigma_{11}$ computed for the two initial
vibrational states roughly coincides.  However, the dynamics and
timescales to relax to the steady-state are clearly sensitive to the
initial vibrational preparation.

\begin{figure}
\includegraphics[width=8cm]{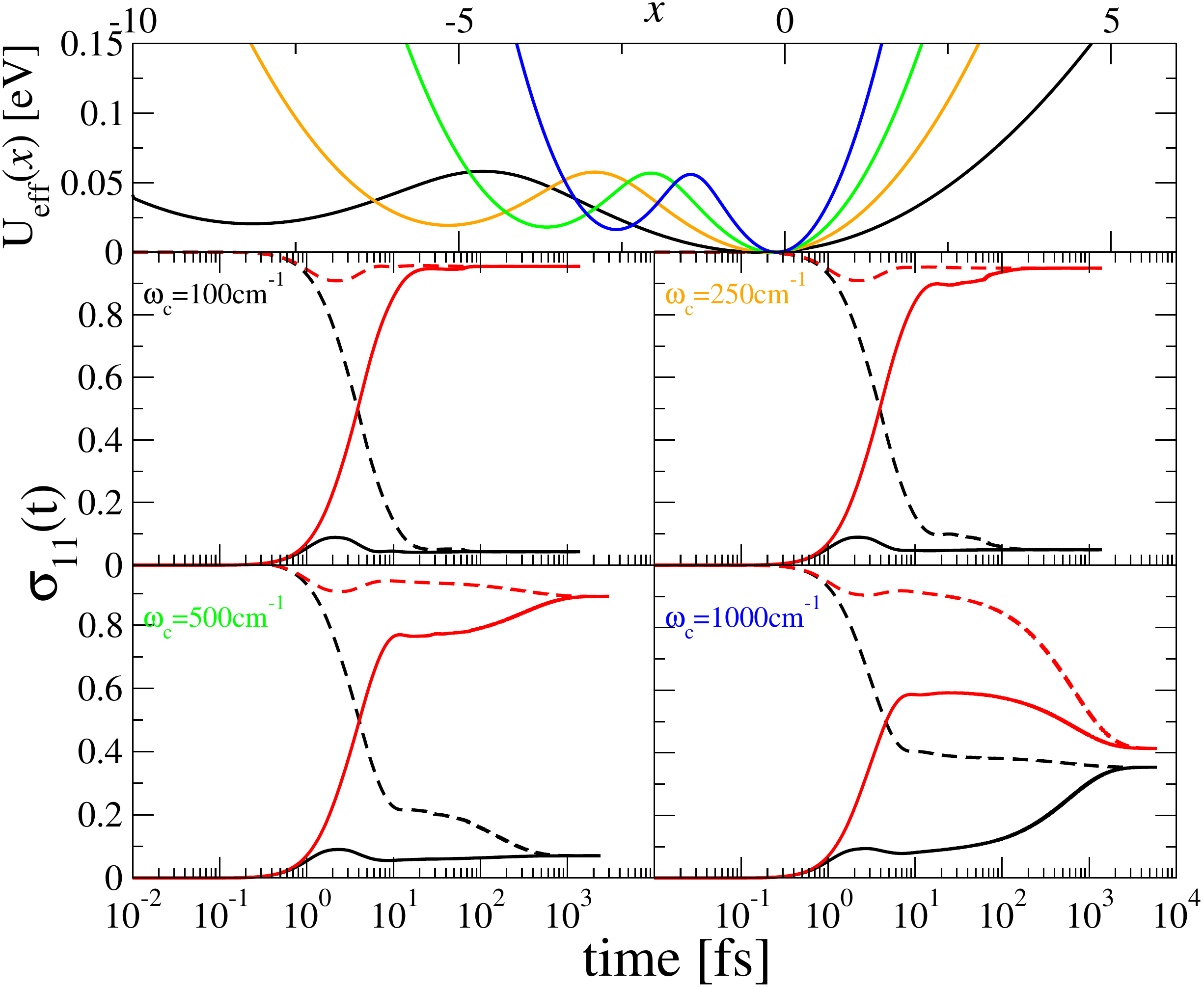}
\caption{Plots of the occupation of the quantum dot $\sigma_{11}(t)$
  for different $\omega_{c}$. The solid black, dashed black, solid
  red, and dashed red curves correspond to $\{n_{d}=0,
  \delta_{\alpha}=0\}$, $\{n_{d}=1, \delta_{\alpha}=0\}$, $\{n_{d}=0,
  \delta_{\alpha}=\sqrt{2\omega_{\alpha}\lambda}\}$, and $\{n_{d}=1,
  \delta_{\alpha}=\sqrt{2\omega_{\alpha}\lambda} \}$,
  respectively. Upper panel shows the adiabatic effective potentials
  for the different values of $\omega_c$.  The model parameters used
  are: $\varepsilon_d=0.5$eV, and $\lambda=4000\mbox{cm}^{-1}$.}
\label{fig:omega_c_dep}
\end{figure}

In Fig.~\ref{fig:omega_c_dep} we plot $\sigma_{11}(t)$ for four
different values of $\omega_c$ and compare the time dependence for
four different initial conditions, corresponding to an initial empty
($n_d=0$) or occupied ($n_d=1$) dot and to $\delta_{\alpha}=0$ or
$\delta_{\alpha}=\sqrt{2\omega_{\alpha}\lambda}$.  For a given
$\delta_{\alpha}$, we find that $\sigma_{11}(t)$ has a unique
steady-state solution regardless of the initial dot occupation.  This
numerically converged result is consistent with the analytical proof
given above.  In contrast, for two different initial vibrational
states, a clear bistability is observed and the RDM decays to a
different steady-state solution depending on the value of
$\delta_{\alpha}$.  We note in passing that a similar phenomenon is
known to occur at equilibrium, for example at $T=0$ in the spin-boson
model~\cite{Wang2008,Weissbook}.

The appearance of a bistability is consistent with predictions based
on a mean field treatment, which is accurate in the adiabatic limit
where $\omega_c \rightarrow 0$~\cite{galperin_non-linear_2008}.  For
all four frequencies studied, the adiabatic effective
potentials~\cite{Kosov2011a} shows two distinct minima (upper panel of
Fig.~\ref{fig:omega_c_dep}), corresponding to two possible stable
configurations~\cite{adiabatic}.  The height of the barrier between
the two minima is independent of the phonon frequency, however, as
clearly evident in the figure, the width of the barrier increases as
$\omega_c$ decreases.  This implies that the tunneling time between
the two configurations also increases as $\omega_c$ decreases, and
thus, the bistability (given by the difference between $\sigma_{11}$
at steady-state for $\delta_{\alpha}=0$ and $\delta_{\alpha}\ne 0$)
would increase with decreasing $\omega_c$, as indeed is the case.  

The transient dynamics and the approach to steady state depends
sensitively on the preparation, in particular, whether the initial
state is close to equilibrium ($\{n_d=0, \delta_{\alpha} =0\}$ and
$\{n_d=1, \delta = \sqrt{2\omega_{\alpha}\lambda} \}$) or far from
equilibrium ($\{n_d = 0,\delta_{alpha}=\sqrt{2\omega_{\alpha}\lambda}
\}$ and $\{n_d=1$, $\delta_{\alpha}=0 \}$).  For small $\omega_c$ we
observe a rapid decay to steady state when the initial preparation is
close to equilibrium, while for the nonequilibrium preparation, the
population of the dot $\sigma_{11}(t)$ decays to steady state on a
time scale $\hbar \Gamma^{-1}$, where $\Gamma$ denotes the maximum of
leads spectral function.  As $\omega_c$ increases, the dynamics
becomes more complex.  In particular, a longer time scale in the
relaxation of $\sigma_{11}(t)$ developes, consistent with the
appearance of a slow tunneling channel between the two configurations.
However, the existence of this channel does not lead to a unique
steady state solution for the populations at a finite value of
$\omega_c$, away from the adiabatic limit.

\begin{figure}
\includegraphics[width=8cm]{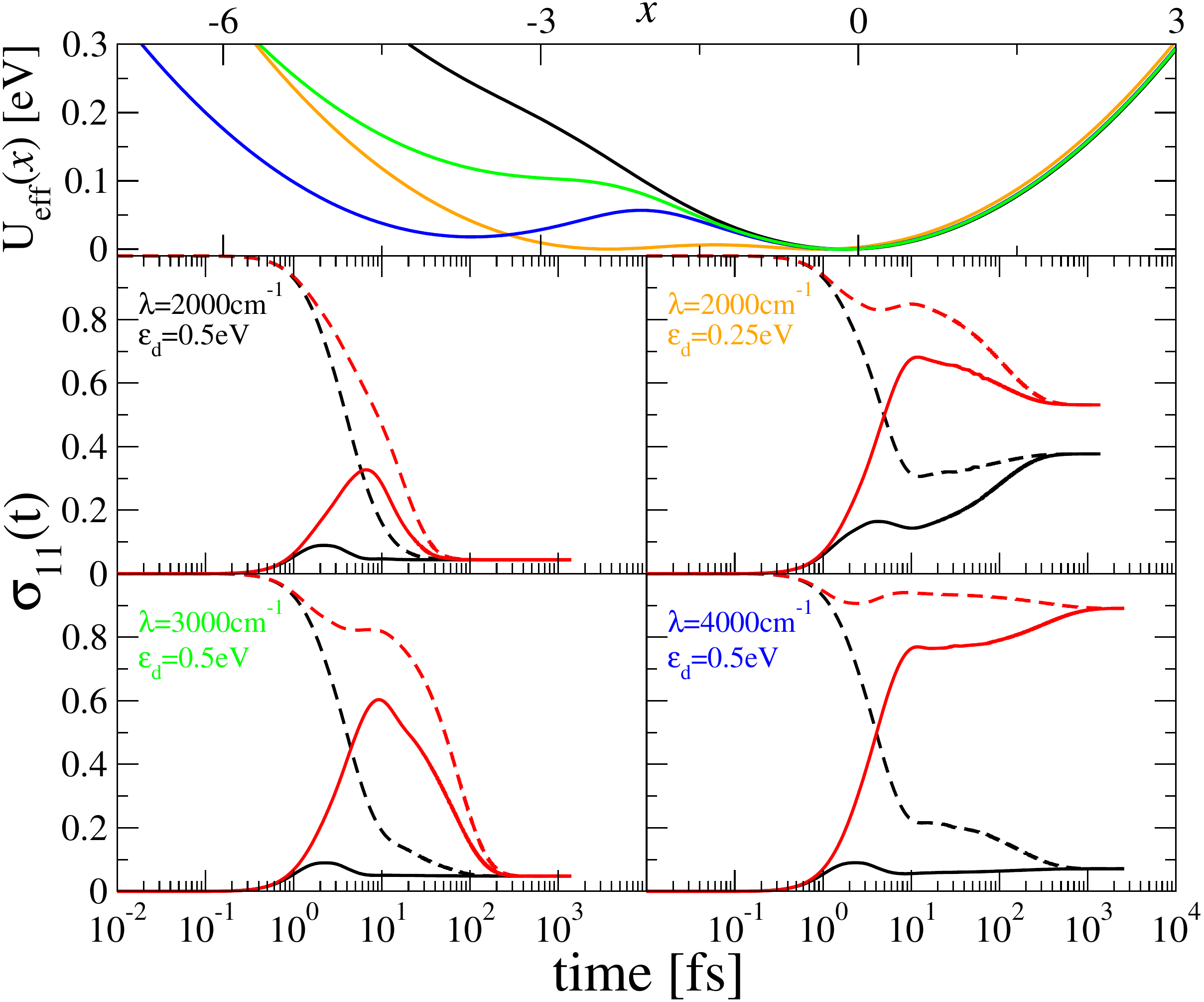}
\caption{Similar to Fig.~\ref{fig:omega_c_dep} but for different
  values of $\lambda$ and $\varepsilon_d$, for
  $\omega_c=500\mbox{cm}^{-1}$. Upper panel shows the adiabatic
  effective potentials for the different values of $\lambda$.}
\label{fig:lambda_dep}
\end{figure}

In Fig.~\ref{fig:lambda_dep} we show $\sigma_{11}(t)$ for different
values of the reorganization energy (polaron shift) $\lambda$ and the
dot energy $\varepsilon_d$. As before, we compare the time dependence
of $\sigma_{11}(t)$ for four different initial conditions.  In the
upper panel we show the corresponding adiabatic effective
potentials~\cite{Kosov2011a} for the four values of $\lambda$.  For
small values of $\lambda$, the bistability clearly disappears (left
panels of Fig~\ref{fig:lambda_dep}). This is consistent with the
adiabatic effective potentials showing a single minimum for $\lambda
\le 3000\mbox{cm}^{-1}$ (in fact, a crude estimate based on a
mean-field approach suggests that below $\lambda=3150\mbox{cm}^{-1}$
the bistability vanishes for the current parameters).  Comparing the
relaxation time for $\lambda=2000\mbox{cm}^{-1}$ and
$\lambda=3000\mbox{cm}^{-1}$, we find that the latter is slower,
particularly for the case of
$\delta_{\alpha}=\sqrt{2\omega_{\alpha}\lambda}$.

When $\lambda$ is further increased to $4000\mbox{cm}^{-1}$
(corresponding to a nearly symmetric case where the polaron shift
equals $\varepsilon_d$) the relaxation time stretches even more and
the system decays to a different steady-state depending on the value
of $\delta_{\alpha}$, again consistent with the appearance of two
stable configurations in the corresponding adiabatic effective
potential (upper panel of Fig.~\ref{fig:lambda_dep}).  While the RDM
shows a distinct bistability, it is interesting to note that this is
not the case for the current through the quantum dot (not shown),
which has a unique steady state for the symmetric case ($\lambda
\approx \varepsilon_d$).

In the upper right panel of Fig.~\ref{fig:lambda_dep} we show results
for the case where $\lambda=2000\mbox{cm}^{-1}$ and
$\varepsilon_d=0.25$eV.  The effective adiabatic potential for this
case clearly shows two distinct minima, however, the barrier is lower
than $\lambda=4000\mbox{cm}^{-1}$ and $\varepsilon_d=0.5$eV.
Comparing the two right panels of Fig.~\ref{fig:lambda_dep}, we find
that as $\lambda$ and $\varepsilon_d$ are decreased simultaneously,
the bistability decreases and the timescale to relax to steady-state
also decreases, consistent with the adiabatic tunneling picture
discussed above.

In this work, we have combined a RDM formalism with the ML-MCTDH-SQR
approach to study the nonequilibrium dynamics of a many-body quantum
system with electron-phonon couplings.  For a generic model, which is
widely used to describe phonon-coupled electron transport in quantum
dots and single-molecule junctions, we showed that the system may
exhibit pronounced bistability.  The analysis reveals that the
bistability increases for decreasing phonon frequency and depends
sensitively on the electron-phonon coupling. This was rationalized in
terms of the timescales of phonon tunneling between the two adiabatic
configurations, the phonon relaxation time, and the electron
relaxation.  Based on the RDM formalism, we proved that the
bistability is associated with different initial phonon preparations
and not with a different initial dot occupations. The present results
are expected to be of relevance for the interpretation of recent
experiments on charge transport in molecular junction, which showed
hysteretic behavior~\cite{Li03,Loertscher06,Liljeroth07}, and may
facilitate further experimental studies in the field of
nanoelectronics.

The authors would like to thank Michael Galperin, Abe Nitzan, and Avi
Schiller for useful discussion.  EYW is grateful to The Center for
Nanoscience and Nanotechnology at Tel Aviv University for a doctoral
fellowship.  HW acknowledges the support from the National Science
Foundation CHE-1012479.  GC is grateful Yad Hanadiv Rothschild
Foundation for the award of a Rothschild Fellowship. MT and ER are
grateful to the Institute of Advanced Studies at the Hebrew University
for the warm hospitality.  This work was supported by the FP7 Marie
Curie IOF project HJSC, by the DFG (SFB 953 and cluster of excellence
EAM), and used resources of the National Energy Research Scientific
Computing Center, which is supported by the Office of Science of the
U.S.  Department of Energy under Contract No. DE-AC02-05CH11231.


\begin{thebibliography}{10}
\bibitem{adiabatic}%
  \BibitemOpen
  \bibinfo {note} {To obtain the adiabatic potential the multiphonon problem
  was represented by a single mode with potential $U(x)=\frac{1}{2}\omega_c
  x^2$. The adiabatic effective potential is then given by the integral over
  the mean force $U_{\mbox{eff}}(x)=U(x)+\int dx \varepsilon^{\prime}(x) n(x)$
  where $\varepsilon(x)=\varepsilon_d+M_c x$ is the dot energy at position $x$,
  and $n(x)=-\frac{i\hbar}{2\pi}\int_{\infty}^{\infty} d\omega
  G^{<}(\omega,x)$, where $G^{<}(\omega,x)$ is the lesser Green function for a
  fixed phonon coordinate obtained by solving the nonequilibrium Green function
  on the Keldysh contour (see Ref. [26]).}%
  \bibAnnoteFile{Stop}{adiabatic}%
\end{thebibliography}
\end{document}